\documentclass[aps,prl,showpacs,twocolumn,superscriptaddress]{revtex4}
\usepackage{bm,color}
\usepackage{graphicx}
\usepackage{amsmath}
\usepackage{color}
\pdfoutput=1

\begin{document}
\title {Theory of Magnetism-Driven Negative Thermal Expansion in Inverse Perovskite Antiferromagnets}
	
\author{Masaya Kobayashi}
\affiliation{Department of Physics and Mathematics, Aoyama Gakuin University, Sagamihara, Kanagawa 229-8558, Japan}
\author{Masahito Mochizuki}
\affiliation{Department of Physics and Mathematics, Aoyama Gakuin University, Sagamihara, Kanagawa 229-8558, Japan}
\affiliation{Department of Applied Physics, Waseda University, Okubo, Shinjuku-ku, Tokyo 169-8555, Japan}
\begin{abstract}
Magnetism-induced negative thermal expansion (NTE) observed in inverse perovskite antiferromagnets ${\rm Mn}_3A{\rm N}$ ($A$=Zn, Ga, etc) is theoretically studied by a classical spin model with competing bond-length-dependent exchange interactions. We numerically reproduce the crystal-volume expansion upon cooling triggered by a non-coplanar antiferromagnetic order and show that the expansion occurs so as to maximize an energy gain of the nearest-neighbor antiferromagnetic interactions. This mechanism is not specific to inverse perovskite magnets and might also be expected in magnets with other crystal structures. We propose other candidate crystal structures that might exhibit NTE through this mechanism.
\end{abstract}
\maketitle

\section{Introduction}
\begin{figure}
\begin{center}
\includegraphics[width=1.0\columnwidth]{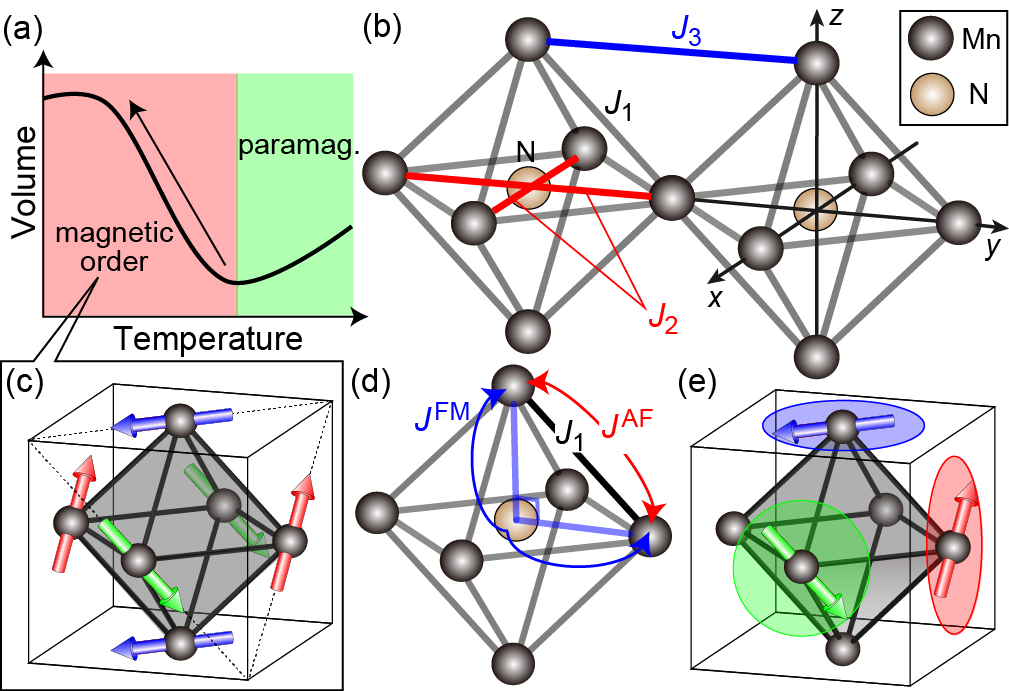}
\caption{(color online). (a) Schematic figure of the negative thermal expansion where the crystal volume shows a pronounced expansion upon cooling at the magnetic transition. (b) Exchange interactions considered for the classical Heisenberg model in Eq.~(\ref{eq:model}). (c) $\Gamma^{5g}$-type antiferromagnetic order in ${\rm Mn}_3A{\rm N}$. (d) Two opposite contributions to the nearest-neighbor exchange interaction $J_1$, i.e., the antiferromagnetic contribution $J^{\rm AF}$ and the ferromagnetic contribution $J^{\rm FM}$. (e) Easy magnetization plane for each Mn sublattice.}
\label{Fig01}
\end{center}
\end{figure}
Nontrivial spin order in frustrated magnets often causes interesting physical phenomena, with useful device functionalities, via coupling to the lattice degrees of freedom. Multiferroic phenomena or magnetoelectric effects in rare-earth perovskite manganites $R$MnO$_3$ ($R$=Tb, Dy, Eu$_{1-x}$Y$_x$) are a typical example of spiral spin order inducing ferroelectric polarization via the inverse effect of Dzyaloshinskii-Moriya interactions~\cite{Kimura03a,Khomskii06,Kimura07,Tokura06a}. Another important example is the magnetovolume effect where the crystal volume shows an abrupt and pronounced change upon a magnetic phase transition~\cite{Takenaka12,Takenaka14}.In general, materials contract in volume as temperature decreases. However, there are some rare examples of respective volume expansion and shrinkage upon cooling and heating [see Fig.~\ref{Fig01}(a)]. This phenomenon is termed negative thermal expansion (NTE)~\cite{ChuCN87,Sleight98,Barrera05,ChenJ15,Takenaka18}. Manganese nitrides ${\rm Mn}_3A{\rm N}$ ($A$=Zn, Ga etc) with an inverse perovskite crystal structure [Fig.~\ref{Fig01}(b)] are a typical class of materials that feature NTE of a magnetic origin~\cite{Bouchaud68,Fruchart71}.

The NTE phenomenon is technically useful for e.g., optical and mechanical parts of devices sensitive to changes in volume and length because composites of negative and positive thermal expansion materials enable control or suppression of changes in length and/or volume. The research field of NTE is rapidly growing recently along with developments of the high-precision devices. This interesting phenomenon is considered to be associated with coupling between a nontrivial spin order shown in Fig.~\ref{Fig01}(c) and the crystal lattice~\cite{Fruchart78,Kaneko87,Iikubo08,Kodama10}. Recently, a microscopic spin model for ${\rm Mn}_3A{\rm N}$ antiferromagnets has been proposed and an origin of the observed noncollinear spin order [see Fig.~\ref{Fig01}(c)] has been clarified using the proposed spin model~\cite{Mochizuki18}. However, the physical mechanism for this magnetism-induced volume expansion has yet to be elucidated.

In this paper, we theoretically study a microscopic mechanism of the magnetism-induced NTE phenomenon for the inverse-perovskite antiferromagnets ${\rm Mn}_3A{\rm N}$ using a classical Heisenberg model, including bond-length-dependent spin exchange interactions. We successfully reproduce the observed crystal-volume expansion upon cooling triggered by a magnetic phase transition to the so-called non-coplanar $\Gamma^{5g}$ antiferromagnetic order. As for its physical mechanism, we clarified that this NTE occurs so as to maximize the energy gain of the nearest-neighbor antiferromagnetic interactions. This finding will help clarify the observed interesting properties of NTE~\cite{Takenaka05,Takenaka06,Takenaka08,Hamada11,HuangR08,SunZH09,SunY07,SunY10,SongXY11} and magnetovolume effects~\cite{Asano08,Takenaka10,WenYC10,Tohei03,WangBS09,Kashima00,WangBS09b} in inverse perovskite antiferromagnets. Moreover, the mechanism we reveal appears to be general and might operate in the NTE, not only of inverse perovskite magnets, but also in magnets with other crystal structures. Some candidate crystal structures that might host magnetism-induced NTE are proposed.

\section{Spin-Lattice Model}
We first examine the spin-lattice coupling in ${\rm Mn}_3A{\rm N}$ via the bond-length dependence of the nearest-neighbor antiferromagnetic exchange $J_1$ [see Fig.~\ref{Fig01}(a)]. We expect that this exchange interaction is composed of two opposite contributions. Namely, the direct Mn-Mn path gives an antiferromagnetic contribution $J^{\rm AF}$($>0$), whereas the indirect 90$^\circ$ Mn-N-Mn path mediated by N ions at the center of the octahedron gives a ferromagnetic contribution $J^{\rm FM}$($<0$) according to the Kanamori-Goodenough rule~\cite{Kanamori59,Kanamori60,Goodenough55,Goodenough58}. These two contributions have different bond-length dependence, which result from distinct distance dependencies of the orbital hybridizations. 

In the atomic limit, the magnitude of the transfer integrals $t_{dd}$ between neighboring $3d$ orbitals is proportional to $\ell^{-5}$, whereas that of the transfer integrals $t_{dp}$ between neighboring $3d$ and $2p$ orbitals is proportional to $\ell^{-7/2}$ with $\ell$ being a distance between the orbitals. These scaling relations can be derived analytically from explicit formula of the $3d$ and $2p$ orbitals, and thus hold universally. On the other hand, these relations might be modified in crystals with periodically aligned atoms because the atomic-orbital picture is no longer valid due to the covalency effects where the orbitals should be described by Wannier functions. However, the first-principles calculations have revealed that the covalency effects on the scaling relations are negligible, and the relations survive even in compounds~\cite{HarrisonTB}. 

The distinct distance dependencies of $t_{dd}$ and $t_{dp}$ give rise to different bond-length dependencies of the two contributions $J^{\rm AF}$ and $J^{\rm FM}$. The antiferromagnetic contribution $J^{\rm AF}$ originates from the second-order perturbation with respect to the transfer integrals $t_{dd}$ between the Mn$3d$ orbitals. This leads to a relation $J^{\rm AF}\propto t_{dd}^2\propto \ell^{-10}$ with $\ell$ being a distance between the Mn ions. Consequently, we obtain the relation,
\begin{eqnarray}
J^{\rm AF} \propto (1+\delta)^{-10} \;\sim\; 1-10\delta.
\label{eqn:JAF}
\end{eqnarray}
Here $1+\delta$$=(\ell_0+\Delta \ell)/\ell_0$ is an elongated or shortened bond length, $\ell_0+\Delta \ell$, normalized to the original bond length $\ell_0$ with $\delta=\Delta \ell/\ell_0$ being the normalized difference of the bond length.

Conversely, the ferromagnetic contribution $J^{\rm FM}$ originates from the fourth-order perturbation with respect to the transfer integrals $t_{dp}$ between the Mn$3d$ and N$2p$ orbitals, resulting in a relation $J^{\rm FM}\propto t_{dp}^4\propto \ell^{-14}$ with $\ell$ being a distance between the Mn and N ions. As a result, we obtain the relation:
\begin{eqnarray}
J^{\rm FM} \propto (1+\delta)^{-14} \;\sim\; 1-14\delta.
\label{eqn:JF}
\end{eqnarray}

On the basis of the above argument, we obtain the following expression of the volume-dependent nearest-neighbor antiferromagnetic interaction $J_1(\delta)$
\begin{eqnarray}
J_1(\delta)= J^{\rm AF}(1-10\delta)-|J^{\rm FM}|(1-14\delta).
\label{eq:Jnn}
\end{eqnarray}
Substituting this expression into the previously proposed spin model for ${\rm Mn}_3A{\rm N}$ in Ref.~\cite{Mochizuki18}, we obtain the following spin-lattice model on the inverse perovskite lattice,
\begin{eqnarray}
\mathcal{H}&=&
 \sum_{<i,\mu;j,\nu>}J_1(\delta_k) \bm S_{i,\mu} \cdot \bm S_{j,\nu}
\nonumber \\
& &+J_2 \sum_{(i,\mu;j,\nu)} \bm S_{i,\mu} \cdot \bm S_{j,\nu}
+J_3 \sum_{\{i,\mu;j,\nu\}} \bm S_{i,\mu} \cdot \bm S_{j,\nu}
\nonumber \\
& &+A \sum_{i, \mu} (\bm S_{i,\mu} \cdot \bm e_\mu)^2
+K_1\sum_k \delta_k^2 - K_2\sum_k \delta_k^3
\label{eq:model}
\end{eqnarray}
where $\bm S_{i,\mu}$ denotes the normalized classical spin vector on the $\mu$th Mn sublattice of the $i$th octhahedron. Here the first term with $J_1(\delta_k)$ denotes the nearest-neighbor antiferromagnetic exchange interactions where $k$ denotes an Mn$_6$N octahedron to which the nearest neighbor bond connecting the two adjacent spins $\bm S_{i,\mu}$ and $\bm S_{j,\nu}$ belongs. The second and third terms with $J_2$ and $J_3$ describe the next-nearest neighbor ferromagnetic exchange interactions where the former and latter respectively correspond to the bonds within an octahedron and between octahedra. The fourth term represents the easy-plane magnetic anisotropy with different easy planes depending on the Mn sublattice [see Fig.~\ref{Fig01}(e)]. For details of the magnetic anisotropies, see Ref.~\cite{Mochizuki18}. The last term denotes the elastic energy which includes, in addition to the harmonic term, a higher-harmonic term proportional to $\delta^3$, which reproduces the usual volume contraction upon cooling.
We adopt $J^{\rm AF}=1$ as the energy units and set $J_2=-0.5$, $J_3=-0.5$, while $J^{\rm FM}$ is taken to be a variable. 

The value of $K_1$ is evaluated to be 3000 so as to reproduce the experimentally observed value of $\bar{\delta}$ at the lowest temperatures. We used the replica exchange Monte-Carlo method to analyze this classical spin model. The spin vectors $\bm S_{i,\mu}$ and the normalized bond length $1+\delta_k$ are updated by the heat-bath method. In the present calculations, we assume isotropic expansions of the Mn$_6$N octahedra with respect to the principal axes of cubic coordinates, because the $\Gamma^{5g}$ type antiferromagnetic order triggering the negative thermal expansion has a cubic symmetry. This assumption has been supported by experiments which indeed observed the isotropic expansion keeping the cubic crystal symmetry~\cite{Takenaka12,Takenaka14}. We also mention that a theoretical analysis based on a localized spin picture was successfully applied to the NTE in invar alloys previously~\cite{Hausch73}.

\section{Results}
\begin{figure}
\begin{center}
\includegraphics[width=1.0\columnwidth]{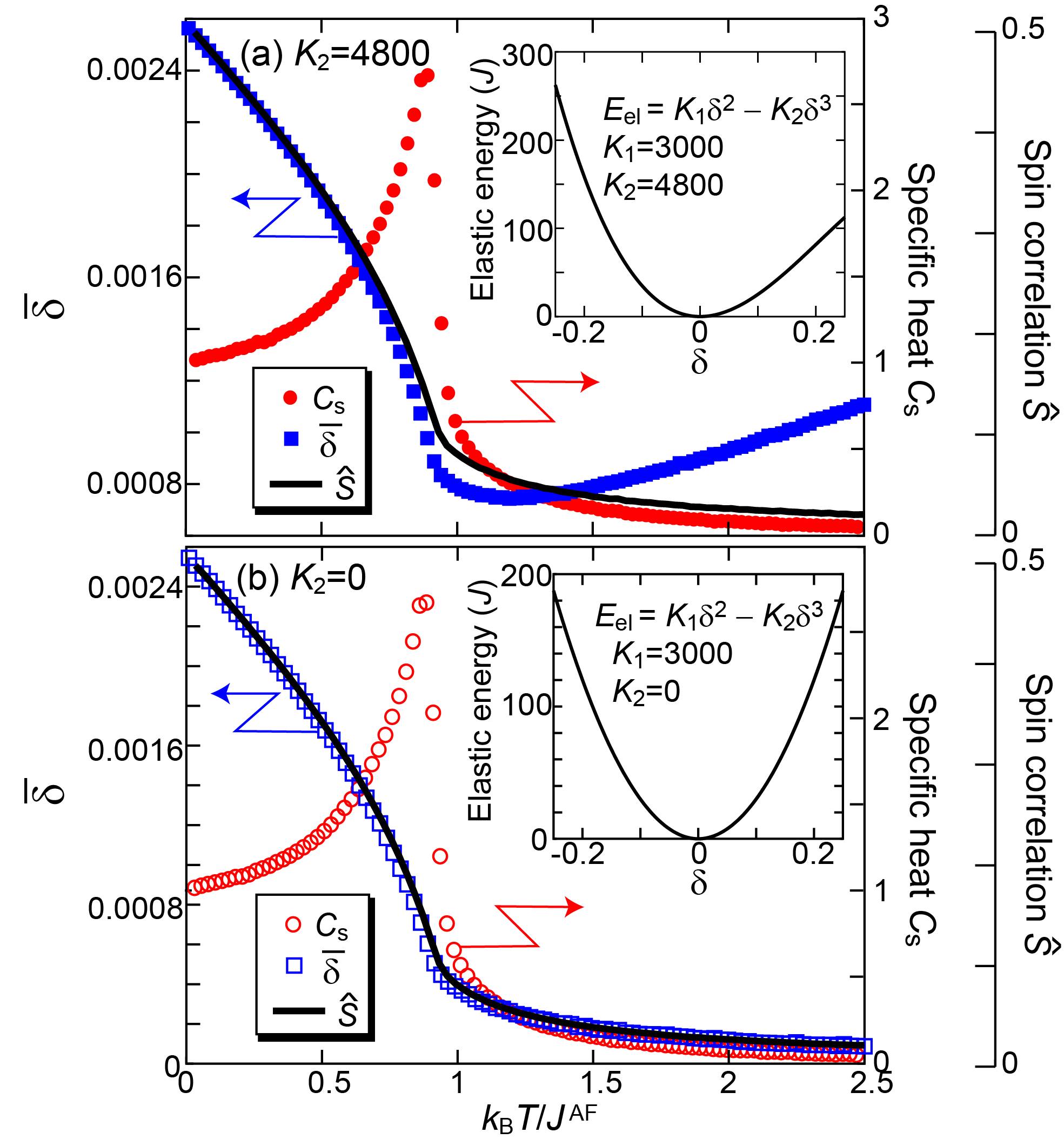}
\caption{(color online). (a) [(b)] Calculated specific heat $C_{\rm s}$, averaged difference of normalized bond length $\bar{\delta}=(1/N)\sum_k \delta_k$, and spin correlation $\hat{S}=(-1/N_{\rm pair})\sum_{<i,\mu;j,\nu>} \langle \bm S_{i,\mu}\cdot\bm S_{j,\nu} \rangle$ as functions of temperature in the presence [absence] of the anharmonic component with $K_2=4800$ [$K_2=0$] in the lattice elastic-energy term. The quantity $\bar{\delta}$ is related to the linearized crystal volume as $\bar{\delta}\propto \left[V(T)/V_0\right]^{1/3}-1$ where $V_0$ is the original volume. Insets show $\delta$ dependence of the lattice elastic energy per octahedron, which shows an anharmonic [a harmonic] behavior for (a) [(b)] with $K_2=0$ [$K_2=4800$].}
\label{Fig02}
\end{center}
\end{figure}
Figure~\ref{Fig02}(a) shows temperature dependence of the specific heat and that of the averaged difference of normalized bond length $\bar{\delta}=(1/N)\sum_k \delta_k$ calculated by considering a finite higher harmonic component $K_2$ for the lattice elastic energy. The value of $K_2$ is taken to be 4800 so as to reproduce the observed conventional volume contraction with decreasing temperature above the magnetic transition point. The sharp peak in the specific heat indicates a magnetic phase transition at $k_{\rm B}T \sim 0.9J^{\rm AF}$. Noticeably the bond length or the crystal volume gradually decreases upon cooling until this magnetic transition point, which is a conventional thermally induced contraction of the crystal volume originating from the higher harmonic component of the elastic energy. Conversely, the crystal volume starts increasing as the temperature decreases right after the magnetic transition, indicating that the NTE induced by the $\Gamma^{5g}$-type antiferromagnetic order is successfully reproduced in our model. 

In Fig.~\ref{Fig02}(a), we also plot calculated thermal averages of the spin correlation,
\begin{eqnarray}
\hat{S}=-\frac{1}{N_{\rm pair}}\sum_{<i,\mu;j,\nu>} \langle \bm S_{i,\mu}\cdot\bm S_{j,\nu} \rangle,
\end{eqnarray}
with $\bm S_{i,\mu}$ and $\bm S_{j,\nu}$ being adjacent spin pairs and $N_{\rm pair}$ being the number of spin pairs summed up, which is calculated in the Monte-Carlo simulations. For the $\Gamma^{5g}$ type magnetic order at $T$=0, this quantity should be 0.5. We find that the volume expansion starts and grows in conjunction with kink and growth of the spin correlation with decreasing temperature, indicating that the negative thermal expansion is indeed driven by the $\Gamma^{5g}$ type antiferromagnetic order.

We also examine the case without the higher harmonic elastic term by setting $K_2=0$ and find that the thermally induced volume change above the transition point vanishes; however, the volume expansion after the magnetic transition is again reproduced. We find that the magnetically induced NTE phenomenon below the magnetic transition temperature can be reproduced well even without the higher harmonic elastic term. Based on this argument, we neglect this term in the calculations for Fig.~\ref{Fig04} to focus on the pure effect of spin-lattice coupling. It is also worth mentioning that the temperature profile of the spin correlation $\hat{S}$ coincides with that of the linearized volume expansion $\bar{\delta}$ almost perfectly, indicating that the volume variation is governed purely by the magnetism in the absence of the higher harmonic elastic term.

\begin{figure}
\begin{center}
\includegraphics[width=1.0\columnwidth]{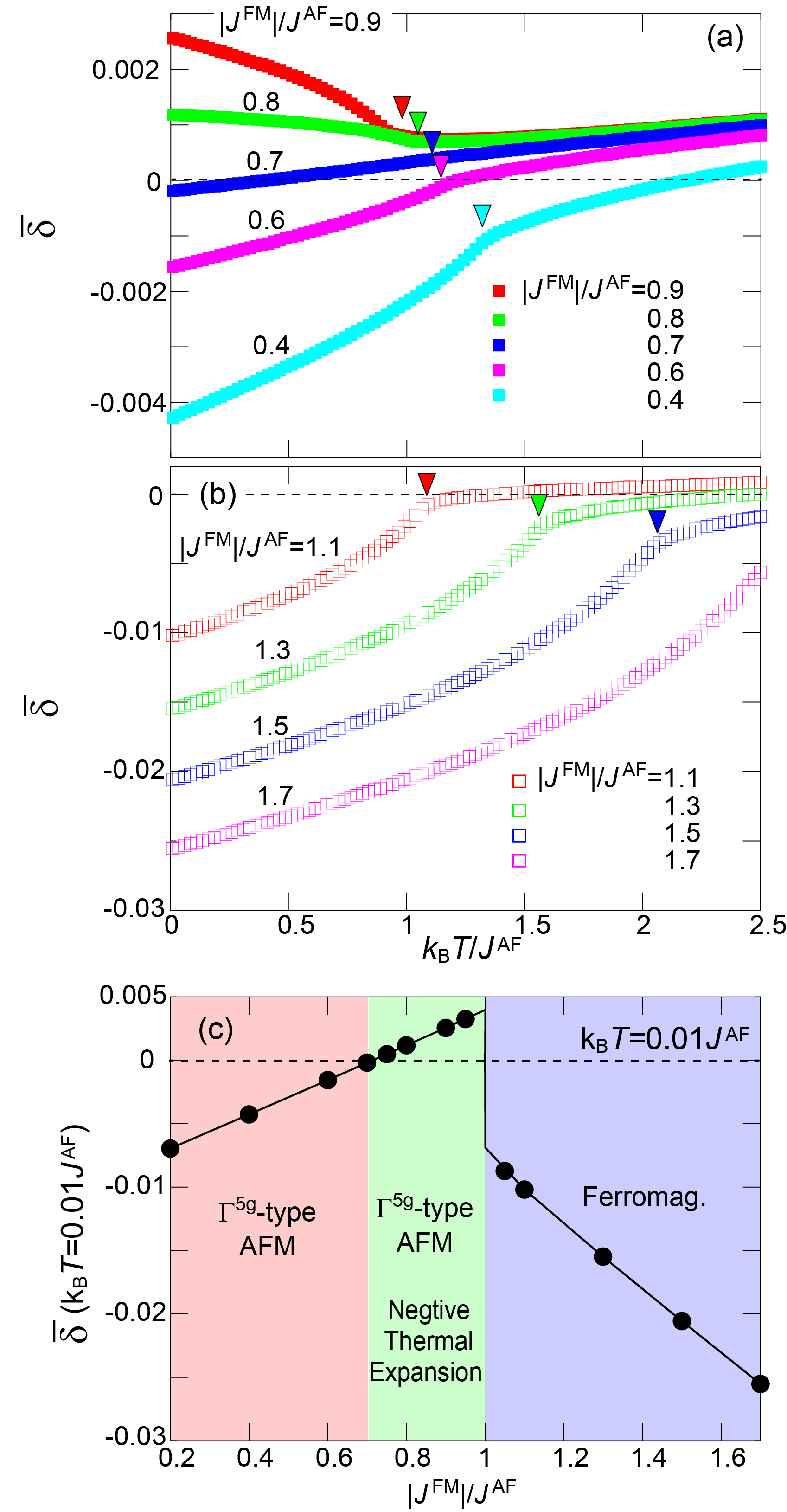}
\caption{(color online). (a) [(b)] Calculated thermal averages of the normalized bond-length difference $\bar{\delta}$ as functions of temperature for several values of $|J^{\rm FM}|/J^{\rm AF}$ ($\le 0.9$) [($\ge 1.1$)]. A positive (negative) value of $\bar{\delta}$ indicates expansion (contraction) of the crystal from its original volume. Upturns after the magnetic transition upon cooling indicates that the NTE is triggered by the magnetic transitions as seen in the cases of $|J^{\rm FM}|/J^{\rm AF}$=0.8 and 0.9. The magnetic transition temperatures are indicated by inverted triangles. (c) Calculated $\bar{\delta}$ at ${\rm k}_{\rm B}T$=0.01$J^{\rm AF}$ as a function of $|J^{\rm FM}|/J^{\rm AF}$, which shows that the $\Gamma^{5g}$-type antiferromagnetic order does not necessarily host the magnetism-driven NTE, but it occurs only in the limited area of the $\Gamma^{5g}$-type antiferromagnetic phase. The NTE occurs in the range $0.7\lesssim|J^{\rm FM}|/J^{\rm AF}\lesssim1$, whereas the $\Gamma^{5g}$-type order takes place in $|J^{\rm FM}|/J^{\rm AF}<1$.}
\label{Fig03}
\end{center}
\end{figure}
In Fig.~\ref{Fig03}(a) and (b), we show calculated temperature dependence of the averaged difference of normalized bond length $\bar{\delta}\equiv(1/N)\sum_k \delta_k$ for several values of $|J^{\rm FM}|/J^{\rm AF}$. Figure~\ref{Fig03}(a) [(b)] shows the data for $|J^{\rm FM}|/J^{\rm AF}<1$ [$|J^{\rm FM}|/J^{\rm AF}>1$] where the $\Gamma^{\rm 5g}$-type antiferromagnetic [ferromagnetic] order takes place at low temperatures. We find that the NTE with increasing $\delta(>0)$ upon cooling is observed in the cases of $|J^{\rm FM}|/J^{\rm AF}=0.8$ and 0.9 where the system is in the $\Gamma^{\rm 5g}$-type antiferromagnetic phase but is located in the vicinity of the phase boundary ($|J^{\rm FM}|/J^{\rm AF}=1$) to the ferromagnetic phase. Conversely, the NTE does not occur in the ferromagnetic phase with $|J^{\rm FM}|/J^{\rm AF}>1$. As shown in Fig.~\ref{Fig03}(c), more detailed calculations revealed that the NTE occurs when $0.7\lesssim|J^{\rm FM}|/J^{\rm AF}\lesssim J^{\rm AF}$ in the $\Gamma^{5g}$-type phase. The extent of the volume expansion is more pronounced for a larger value of $|J^{\rm FM}|/J^{\rm AF}$ within this phase. On the contrary, when the $|J^{\rm FM}|/J^{\rm AF}$ exceeds unity and the system enters the ferromagnetic phase, the NTE suddenly vanishes and the crystal volume contracts more noticeably for a larger value of $|J^{\rm FM}|/J^{\rm AF}$ in the ferromagnetic phase.

According to Fig.~\ref{Fig03}(c), we find that there are lower and higher threshold values of $|J^{\rm FM}|/J^{\rm AF}$ for occurrence of the NTE at $\sim0.7$ and $\sim1$. This limited range of $|J^{\rm FM}|/J^{\rm AF}$ can be understood quantitatively. The fact that the NTE is driven by the magnetic transition to the $\Gamma^{5g}$-type antiferromagnetic order means that the crystal-volume expansion further decreases the magnetic energy via enhancing the antiferromagnetic coupling $J_1(\delta)$. Namely, the coupling $J_1$ is expected to increase for a positive $\delta$. This situation is indeed realized when the coefficient of $\delta$-linear term in Eq.~(\ref{eq:Jnn}), $14|J^{\rm FM}|-10J^{\rm AF}$, is positive, which leads to $|J^{\rm FM}|/J^{\rm AF}>10/14\sim0.714$. In addition, the condition $J^{\rm AF}-|J^{\rm FM}|>0$ is required such that a transition to the $\Gamma^{5g}$-type phase occurs to trigger the NTE even when the system is not expanded ($\delta$=0). Accordingly, we obtain the following condition for occurrence of the NTE upon the magnetic transition to the $\Gamma^{5g}$-type antiferromagnetic order:
\begin{eqnarray}
0.714<\frac{|J^{\rm FM}|}{J^{\rm AF}}<1.
\label{eq:condition}
\end{eqnarray}

\begin{figure}
\begin{center}
\includegraphics[width=1.0\columnwidth]{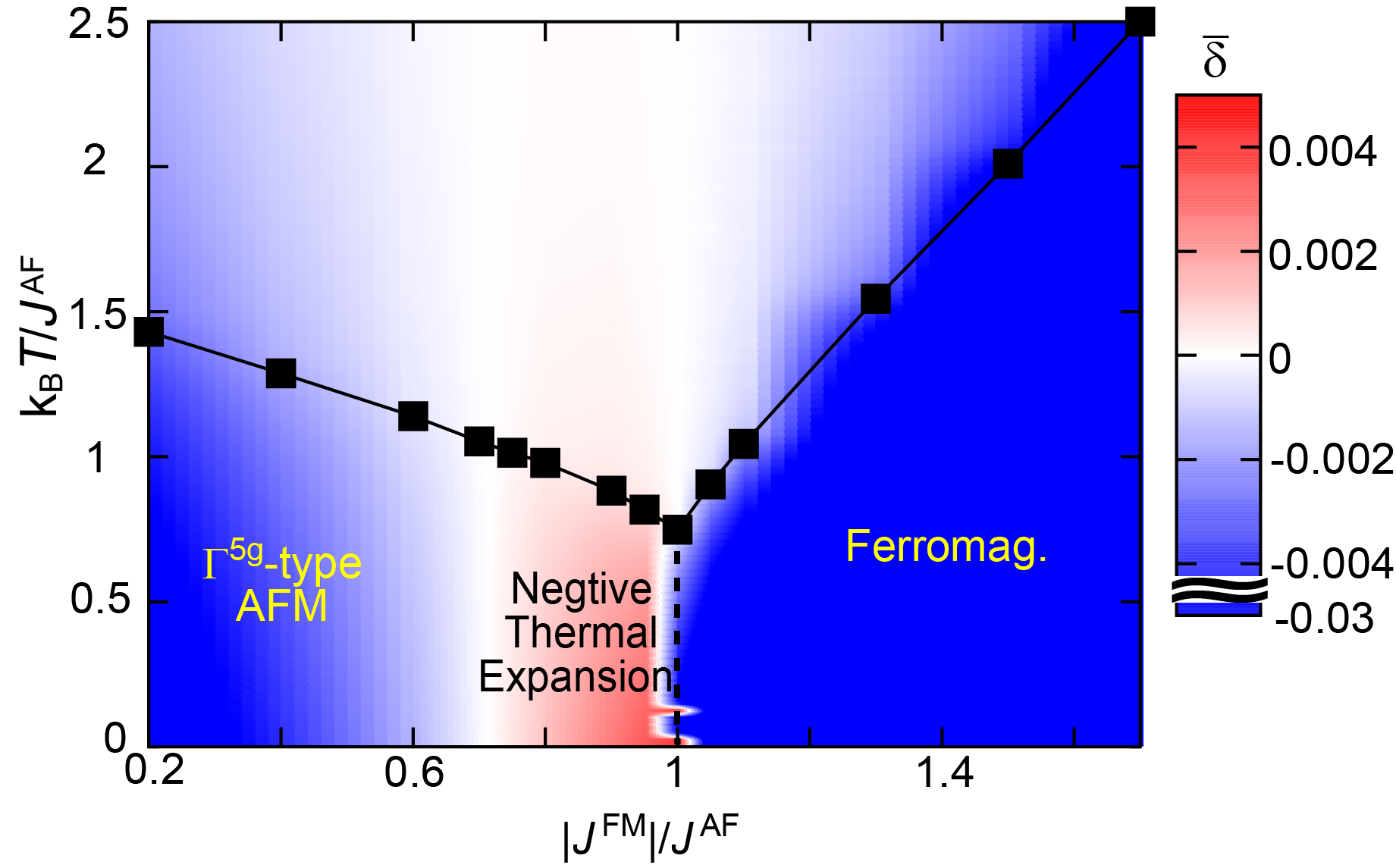}
\caption{(color online). Phase diagram of the spin model in Eq.~(\ref{eq:model}) and color map of the averaged difference of normalized bond length $\bar{\delta}$ in plane of temperature $k_{\rm B}T/J^{\rm AF}$ and the ratio $|J^{\rm FM}|/J^{\rm AF}$, which indicates the occurrence of NTE in the limited area of the $\Gamma^{5g}$-type antiferromagnetic phase with $0.7\lesssim|J^{\rm FM}|/J^{\rm AF}\lesssim1$.}
\label{Fig04}
\end{center}
\end{figure}
Figure~\ref{Fig04} displays a theoretical phase diagram of the spin model in Eq.~(\ref{eq:model}) and a color map of the normalized difference of bond length $\delta$ in plane of temperature and the ratio $|J^{\rm FM}|/J^{\rm AF}$, which clearly shows that the positive $\bar{\delta}$ appears below the transition temperatures to the $\Gamma^{5g}$-type antiferromagnetic phase in the limited range of $0.7 \lesssim |J^{\rm FM}|/J^{\rm AF} \lesssim 1$. As already mentioned above, the higher harmonic elastic term is neglected for the calculations to extract the pure effect of the spin-lattice coupling.

Notably, our spin model produces the second-order phase transition where the bond-length difference $\delta$ continuously increases right below the transition point upon cooling, whereas the experimentally observed phase transition has an intrinsic strong first order nature. This apparent inconsistency can be attributed to our theoretical treatment based on a pure spin model with the presumption of quenched orbital degrees of freedom. We assume an orbital pattern determined by the crystal field from the $A$ ions and that from the N ions and consider the magnetic anisotropies under this orbital pattern. However, in real materials, the orbitals thermally fluctuate at higher temperatures and become ordered at lower temperatures. Hence, the orbitals are also the order parameter of the present system, and mutual coupling of spins and orbitals will result in the strong first order phase transition. A more elaborate theoretical study incorporating the orbital degrees of freedom is left for future study. However, the essential physics of the magnetically induced NTE phenomenon in the inverse perovskite antiferromagnets has been clarified in the present study.

\section{Conclusion and Discussion}
\begin{figure}
\begin{center}
\includegraphics[width=1.0\columnwidth]{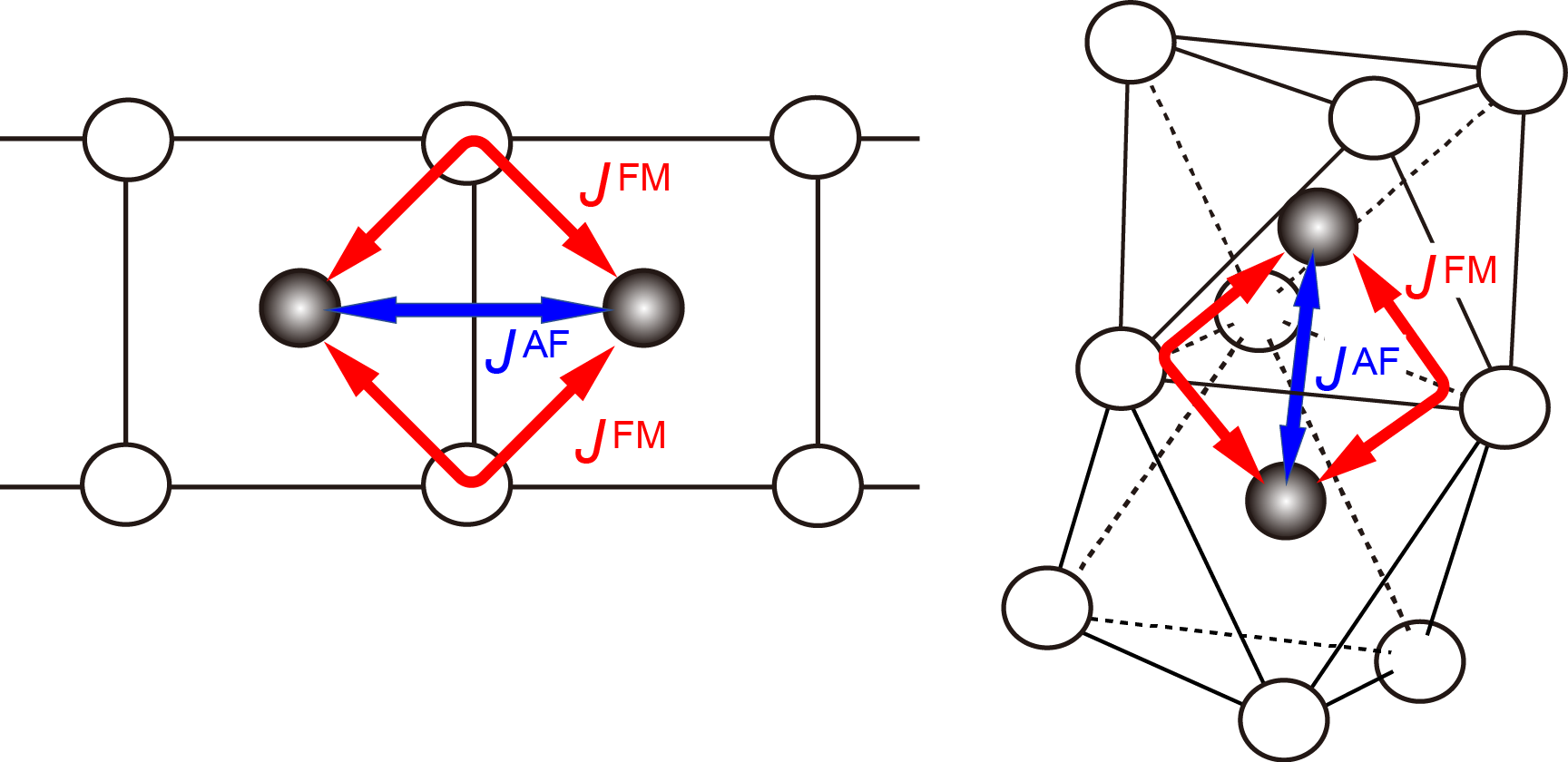}
\caption{(color online). Crystal structures possibly hosting the NTE phenomenon triggered by the antiferromagnetic order, in which keen competition between antiferromagnetic and ferromagnetic contributions to the nearest-neighbor exchange interactions is realized.}
\label{Fig05}
\end{center}
\end{figure}
In conclusion, we have theoretically investigated the experimentally observed crystal-volume expansion upon cooling in the inverse perovskite ${\rm Mn}_3A{\rm N}$. Our classical spin model with the spin-lattice coupling has successfully reproduced the NTE phenomenon triggered by the non-coplanar $\Gamma^{\rm 5g}$ type antiferromagnetic order. We have revealed that the volume expansion or the bond elongation enhances the nearest-neighbor antiferromagnetic coupling $J_1$ in the inverse perovskite structure because the two opposing contributions constituting this coupling have different bond-length dependencies. The antiferromagnetic contribution $J^{\rm AF}$ from the direct Mn-Mn path, which is governed by the second-order perturbation processes of direct $d$-$d$ electron transfers, is weakly suppressed. On the other hand, the ferromagnetic contribution $J^{\rm FM}$ from the 90$^\circ$ Mn-N-Mn path, which is governed by the fourth-order perturbation processes of $d$-$p$ electron transfers, is suppressed significantly. Consequently, the crystal volume and the bond length tend to contract when the magnetic transition takes place so as to increase the antiferromagnetic coupling $J_1$ and the associated energy gain. We note that this mechanism is not specific to the inverse perovskites but might be expected in other crystal structures. Namely, antiferromagnets in which the antiferromagnetic contribution from direct exchange paths and the ferromagnetic contribution from indirect 90$^\circ$ paths severely compete are candidates that host NTE by this mechanism [Fig.~\ref{Fig05}].

We believe that this prediction will be a useful guide to search for new NTE materials because there exists no reliable strategy to seek magnetism-driven NTE materials at present. However, it should be noted that the existence of this competition is not a sufficient condition but a necessary condition for emergence of the magnetism-driven NTE. As we argued clearly, the emergence of $\Gamma^{\rm 5g}$ antiferromagnetic order does not necessarily induce the NTE, but the NTE occurs when the ratio $|J^{\rm FM}|/J^{\rm AF}$ is within a range of $0.714 < |J^{\rm FM}|/J^{\rm AF} < 1$ (see also Fig.~\ref{Fig04}). Because the value of $|J^{\rm FM}|/J^{\rm AF}$ varies depending on materials, it is not easy to discuss the sufficient condition for the NTE. However, we can propose systems that tend to have a large $|J^{\rm FM}|/J^{\rm AF}$, which will be helpful for searching and designing of new NTE materials. In general, the direct antiferromagnetic exchange $J^{\rm AF}$ is given by
\begin{eqnarray}
J^{\rm AF} =\frac{4t_{dd}^2}{U}.
\end{eqnarray}
On the other hand, the 90$^\circ$-bond ferromagnetic exchange $J^{\rm FM}$ is given by 
\begin{eqnarray}
J^{\rm FM} =-\frac{4t_{dp}^4 J_{\rm H}}{\Delta^2 U^2}.
\end{eqnarray}
Accordingly, their ratio becomes
\begin{eqnarray}
|J^{\rm FM}|/J^{\rm AF} = \frac{t_{dp}^4 J_{\rm H}}{t_{dd}^2 \Delta^2 U}.
\end{eqnarray}
Here $t_{dd}$ and $t_{dp}$ represent a transfer integral between neighboring transition-metal $d$ orbitals and that between the transition-metal $d$ and ligand $p$ orbitals, respectively. The symbols $U$ and $J_{\rm H}$ represent strengths of the averaged Coulomb interaction and the Hund's-rule coupling within the $d$ orbitals, respectively, while $\Delta$ denotes the charge transfer energy, i.e., the energy-level difference between the $d$ and $p$ orbitals. In the expression of the ratio $|J^{\rm FM}|/J^{\rm AF}$, we find that larger $t_{dp}$ and $J_{\rm H}$ as well as a smaller $\Delta$ tend to give a larger $|J^{\rm FM}|/J^{\rm AF}$. Note that $t_{dp}$ tends to be large in the $e_g$-orbital compounds with $\sigma$ bonding between the $d$ and $p$ orbitals, while $\Delta$ tends to be small for heavy transition-metal ions. Besides, the Hund's-rule coupling $J_{\rm H}$ is active in partially filled $d$-orbital systems. As a result, we can suggest that transition-metal compounds having 90$^\circ$ $M$-$L$-$M$ bonds with $M$=Mn, Fe, Co tend to have a large $|J^{\rm FM}|/J^{\rm AF}$, and thus are promising for emergence of the magnetism-driven NTE.

\section{Acknowledgment}
This work was supported by JSPS KAKENHI (Grant No. 17H02924), Waseda University Grant for Special Research Projects (Project Nos. 2017S-101, 2018K-257), and JST PRESTO (Grant No. JPMJPR132A). We thank D. Yamamoto for fruitful discussions. 

\end{document}